\documentstyle[12pt,epsfig]{article}
\begin{document}

\date{}

\title{\Large \bf Effect of Friedel oscillations on Capacity of
Tunnel Junction}

\author{ Posvyanskii D.V.
\\\it Institute of
Radioengineering and Electronics \\\it of Russian Academy of
Sciences\\\it 11 Mokhovaya, Moscow, 103907, Russia\\\it} \maketitle

\textwidth 5in
\begin{center}
{\bf Abstract}
\end{center}
\vskip 0.2in
\hskip 0.4in \begin{minipage}[b]{5in} \parindent 0.cm \textwidth 5in
Effect of long range charge density oscillations on capacity of tunnel structure is
studied  within random phase approximation (RPA). Using  this
approximation we obtain expressions for the short and the long distance behavior of the self-consistent
screened potential. We demonstrate that long range  density oscillation, commonly referred to as Friedel oscillations,
leads to decreasing  of total electrostatic  capacity. Particular emphasis has been
placed on influence of an external magnetic field is applied perpendicular to the barrier plane on the capacity of a structure.
It is shown that increasing of magnetic field implies an increase of quantum correction to capacity due to Friedel
oscillations.
\end{minipage}
\textwidth 5.8in

\section{Introduction}
Response of tunneling structures on an external dc bias plays an
important role in a number of tunneling phenomena, such as
zero-bias anomaly and Coulomb blockade. Besides only the fundamental physical
interest, it can have a pronounced effect on performances of electronic devices, based on
semiconductor tunnel junctions.\\
It is well known that the presence of a potential barrier leads to the long-range oscillation of electron
density, which are commonly referred to as the Friedel oscillations. In the contrast to the
three-dimensional case where the electron density oscillations around an impurity decays as $\frac{\displaystyle \cos(2k_{\tt F}\bf{r} )}{\displaystyle\bf
r^3}$ (where $\bf r$ is the distance from the impurity and $k_{\tt F}$ is Fermi wave vector) in the case of metal surface this oscillations
decays only as $\frac{\displaystyle \sin(2k_{\tt F}z)}{\displaystyle
z^2}$, here $z$ is the distance from the surface \cite{1}. In \cite {2} was shown that in one 1D tunneling structure,
charge density  oscillations decay only as $\frac{\displaystyle \cos(2k_{\tt F}z)}{\displaystyle
z}$ and this induces a singularity in differential conductance. This singularity originates from the electron scattering on
the Friedel oscillations.\\
It is well-known that the motions of three-dimensions electrons in an external magnetic
field are quasi one-dimensional and it is possible to expect that
in a strong fields the long-range behaviour of electron density  will be similar to those of
one-dimensional systems.\\
In this paper we investigate effect of Friedel oscillations on capacity  of
barrier structure, it will be shown, that long-range electron density oscillations lead to decreasing of electrostatic  capacity  of the
structure  and this effect plays more important role with increasing external magnetic
field, which is perpendicular to the barrier plane.\\
Formally, the goal is to find a self-consistent distribution of a total electrostatic potential
$\Phi({\bf r})$, since  the knowledge of this potential is sufficient to calculate
capacity of a structure. Self-consistent potential is calculated within RPA approximation, using this approach
we obtained the short- and the long-distance behaviour of electrostatic potential.
\section{General formulation}
Here we briefly summarize the standard self-consistent analysis, which
allows to write an expression for potential $\Phi({\bf r})$ .
In linear response theory the induced charge density is given by
\begin{equation}
\varrho({\bf r})=\int_{-\infty}^{\infty}{\bf dr'}\Pi({\bf r,r'})\Phi({\bf r'})
\end{equation}
here the polarization kernel $\Pi(\bf{r,r'})$ is given by
\begin{equation}
\Pi({\bf r,r'})=\sum_{\bf{k,k'}}\frac{f_{{\bf k'}}-f_{\bf k}}{E_{{\bf k'}}-E_{\bf k} +i0}\\
\Psi_{\bf k}^*({\bf r'})\Psi_{\bf k'}({\bf r'})\Psi_{{\bf k'}}^*({\bf r})\Psi_{\bf k}({\bf r}).
\end{equation}
Hence $f_k$ is the Fermi distribution function (below we will assume that the temperature iz zero), $\Psi_{{\bf k}}$ is wave functions.\\
The potential $\Phi({\bf r})$ is also
related to the charge density by Poisson's equation which can be
written as
$$
\Phi({\bf r})=-4\pi\int G({\bf r,r'})\varrho({\bf r'}),
$$
here $G({\bf r,r''})$ is a Green function of Laplas equation.\\
Combining (1) with (2) gives us an integral equation for
electrostatic potential

\begin{equation}
\Phi({\bf r})=\int {\bf dr'}F({\bf r,r'})\Phi({\bf r'}),
\end{equation}
here
$$
F({\bf r,r'})=\int {\bf dr''}G({\bf r,r''}) \Pi({\bf r'',r'}).
$$
as it was mentioned in introduction, the goal of the present work is to calculate the  the self-consistent
response of tunnel structure on external bias, because formally the knowledge of potential is sufficient
to calculate capacity of a structure.\\
The model we use for a tunnel structure consists two semiconductor half-spaces $z<-d/2$ and
$z>d/2$, separated by a potential barrier of width $d$. A uniform compensating positive background extending through
semiconductors, so electrostatic potential $\Phi({\bf r})$ doesn't
depend on lateral (x-y) coordinates.
Below we will assume that:\\
(i)$\quad$  transparency of the barrier is zero, so
electrostatic potential is a linear function inside a potential barrier.\\
(ii)$\quad$ $\Phi(z)$ is an odd function in z direction.\\
This assumptions allow to transform the task to the problem of an electrical field $E$ penetration
at the surface of semiconductor \cite{3},
\cite{4}. However in the difference from the case of surface of semiconductor, where $E$ had  a fixed value, in our case
the electrical field depends on external dc bias $V$ as well as induced electrostatic potential.
The total electrical field $E$ in the barrier structure can be written as a sum of two terms $E(z)=E_{\tt ext}(z)+E_{\tt
ind}(z)$, where $E_{\tt ext}(z)$ is discontinous function which is equal to a constant $E_{\tt ext}$ inside the barrier and zero in
semiconductor leaders,  $E_{\tt
ind}(z)$ is the field induced by electron systems and it is discontinous function too. The fact, that a
total electrical field should be continues function along the
structure, bring us to the boundary condition for $E_{\tt ind}$ at
$(z=d/2+0)$
\begin{equation}
E \left(\frac{d}{2}\right)=\frac{2\Phi(d/2) + V}{d}.
\end{equation}
It was mentioned that self-consistent potential is calculated
within RPA approximation, which is valid if $r_{\tt s}\ll 1$.
Where
$$
r_{\tt s} = \left(\frac{9\pi}{4}\right)^{\frac{1}{3}}\frac{1}{k_{\tt F}a_{\tt
B}},
$$
here $a_{\tt B}$ is Bohr radius
In this study we ignore considerations of exchange and correlation interactions between electrons.

\section{Self-consistent potential of tunnel structure}
In this section, we present calculations of the self-consistent
potential for a tunneling structure.\\
{\bf The case of zero magnetic field}\\
In this case electrons are described by a normalized eigenfunction
$$
\Psi(r) = \frac{1}{2\pi}e^{i\bf k_{||}r_{||}}\psi_{k_z}(z) ,
$$
where $ \bf k_{\tt ||}$ and $\bf r_{\tt ||}$ are two-dimensional vectors laying in the x-y plane parallel to the barrier
layers and z-dependent part of wave function is
$\psi_{k_z}=\displaystyle \frac{1}{\sqrt{2\pi}}\sin{k_zz}$. On substituting this wave functions in Eq.(3)
we obtain
\begin{equation}
\Phi (z)=
\begin {array}{l}
\displaystyle\int_{d/2}^{\infty}dz'[\Lambda(z-z')+\Lambda(z+z'-d)]\Phi(z')dz'
-\\
\displaystyle\int_{d/2}^\infty K(z,z')\Phi(z')dz'
\end {array}
\quad z>d/2
\end{equation}
\begin{equation}
\Phi''(z)=0 \quad |z|<d/2
\end{equation}
here
$$
\Lambda_q(z)=4\pi e^2 \int_{-\infty}^{\infty}\frac{dk_{||}}{(2\pi)^2}
\int_{0}^{\infty} \frac{dkdk'}{(2\pi)^2} \\
\left(
 \frac{e^{i(k-k')z} +c.c.}{(k-k')^2+q^2} \\
\right)
\frac{ f_{{k',k_{||}+q}}-f_{{k,k_{||}}} }{E_{k',k_{||}+q}-E_{k,k_{||}}+i0}
$$
$$
\begin {array}{l}
K_q(z,z')= 4\pi e^2 \displaystyle \int_{-\infty}^{\infty} \frac{dk_{||}}{(2\pi)^2}\int_{0}^{\infty}
\frac{dkdk'}{(2\pi)^2}
\displaystyle \frac{f_{{k',k_{||}+q}}-f_{{k,k_{||}}} }{E_{k',k_{||}+q}-E_{
k}+i0}\\
\begin {array}{c}
\left( \displaystyle \frac {e^{i(k+k')z}e^{-i(k-k')z'}+ e^{i(k+k')z}e^{i(k-
k')z'}+\tt{c.c.}}{(k+k')^2+q^2}
+
\displaystyle \frac {e^{i(k-k')z}e^{i(k+k')z'}+ e^{i(k-k')z}e^{-i(k+k')z'}+\tt{c.c.}}{(k-
k')^2+q^2}\right),
\end{array}
\end{array}
$$
here $q=(q_x,q_y,0)$ is 2D wave vector in the plane of barrier. As far as $\Phi$ doesn't depend on lateral coordinates,
we will consider the case $q=0$.\\
Making Fourier transforming Eq.(5) with respect to the
variable and some algebra we obtain
\begin{equation}
\Phi(\zeta)=-\frac{1}{2\pi} \int_0^{\infty} d\eta \int_{-\infty}^\infty dq_z
\frac{e^{iq_z\zeta}}{q_z^2\epsilon(q_z)}K(q_z,\eta)\Phi(\eta) \qquad \zeta >0,
\end{equation}
here  $ \zeta=z-d/2 $, $\eta = z'-d/2$, $\epsilon(q_{\tt z},0)$ is the Lindhard dielectric function
$$
\epsilon (q_z,q)= 1 - \displaystyle\int_{-\infty}^{\infty}dze^{-iq_zz}\Lambda_q(z),
$$
$$
 K_q(\zeta,q_z)=\displaystyle \int_{-\infty}^{\infty}d\eta e^{-iq_z\eta}\tilde
K_q(\zeta,\eta).
$$
Integral equation (7) can be considered as a generalization of
well-known Shafranov's equation \cite{5} to the case of non-uniform electron
gas.\\
After integrating over $k_{\tt {||}}$ and $k$, we obtain the
following expression for the integral kernel
\begin{equation}
K(q_z,\eta)=-\frac{ 2{k_{\small \tt TF}}^2k_{\tt F}}{\displaystyle |q_z|}
\left\{
\begin{array}{l}
\left(1-\frac{\displaystyle |q_z|^2}{\displaystyle 4{k_{\tt F}}^2}\right)
({\tt Ci}|2k_{\tt F}\eta+q_z\eta| - {\tt Ci}|2k_{\tt F}\eta-q_z\eta|) \\
+\displaystyle \frac{q_z}{2k_{\tt F}^2\eta}\sin{2k_{\tt F}\eta}\cos{q_z\eta} + \frac{\displaystyle
\sin{2k_{\tt F}\eta}\sin{q_z\eta}}{\displaystyle 2k_{\tt F}^2\eta^2}-\displaystyle\frac{\cos2k_{\tt F}\eta \sin q_z\eta}{k_{\tt F}\eta} ,
\end{array}
\right.
\end{equation}
here $k_{\tt TF}$ is Thomse-Fermi wave vector.
In the long wavelength limit $(q_z \ll k_{\tt F})$ RPA dielectric function  and integral kernel $K$can be
simplified as
$$
\epsilon(q_z) = 1+ \frac{k_{\tt TF}^2}{{q_z}^2}
$$
$$\
K(q_z,\eta)=-{2{k_{\small\rm TF}^2}}\frac{\sin(2k_{\tt F}\eta)}
{2k_{\tt F}\eta}\cos q_z\eta.
$$
In this limit the integral over $q_z$ in (7) is done by closing the
integration contour in the upper half plane and picking up the
pole at $q_z=ik_{\tt TF}$ and Eq.(7) can be transformed to
\begin{equation}
\Phi'' - k_{\tt TF}^2(1-\frac{\sin(2k_{\tt F}\zeta)}{2k_{\tt F}\zeta})\Phi=0
\end{equation}
with boundary conditions
\begin{equation}
\left. {\frac{d\Phi}{d\zeta}}\right|_{\zeta=0} = {E}\qquad
\left. {\frac{d\Phi}{d\zeta}}\right|_{\zeta=\infty} = 0
\end{equation}
In deriving  of Eq.(9) we used the following expression
$$
\frac{d^2}{dx^2}e^{\displaystyle -a|x|} = -2a\delta(x)e^{\displaystyle -a|x|} + a^2 e^{\displaystyle -a|x|}
$$
Here we emphasize that under our assumption (i) electron density must be zero at interface semiconductor-potential barrier $\zeta = 0$
and as it easy to see that solution of Eq.(9) satisfies this condition. It describes the behaviour of  the screened Coulomb potential
if $\zeta \sim \lambda_{\tt TF}$ and it exponentially decays at large distances $\zeta \gg \lambda_{\tt TF}$.\\
Here we would like to repeat that Eq.(9) was obtained with account
of  pole in $\epsilon$, however apart from the pole, the dielectric function has branch point at $q_z=2k_{\tt F}$ and this singularity
generates long-range oscillations in charge density (Friedel oscillations). At large distances $\zeta \gg \lambda_{\tt F}$
this oscillations plays the dominant role in distribution of induced charge. Asymptotic solution of  Eq.(7) can be presented as
$$
\Phi(\zeta)\sim \frac{\displaystyle k_{\tt TF}}{\displaystyle k_{\tt F}}\frac{\displaystyle
\sin{2k_{\tt F}\zeta}}{\displaystyle {(2k_{\tt F}\zeta)}^2}
$$
Thus behaviour of electrostatic potential across the structure is
\begin{equation}
\Phi(\zeta) = \left\{
\begin{array}{lc}
\Phi(0)e^{\displaystyle -k_{\tt TF}\int_0^\zeta dx \sqrt{1-
\frac{\sin{2k_{\tt F}x}}{2k_{\tt F}x }}}  &\zeta<\zeta_c \\
\Phi(0)\frac{\displaystyle k_{\tt TF}}{\displaystyle k_{\tt F}}\frac{\displaystyle
\sin{2k_{\tt F}\zeta}}{\displaystyle {(2k_{\tt F}\zeta)}^2} + o(k_{\tt TF}/{k_{\tt F}})&
\zeta>\zeta_c\\
\end{array}
\right.
\end{equation}
According to \cite{4} parameter $\zeta_c$ can be estimated as
$$
z_c = \lambda_{\tt TF}(\ln{\frac{k_{\tt F}}{k_{\tt TF}}} + \alpha),
$$
where $\alpha > 1$.

{\bf The case of finite magnetic fields}\\
In this section we consider the response of barrier structure on external bias in the presence of magnetic fields
perpendicular to the planes of the barrier. Under the Landau  gauge
with vector potential $A = (-Hy,0,0)$ the wave functions and
corresponding energy levels can be specified by the set of quantum
numbers $(n,k_{\tt z})$ as
$$ \Psi_\alpha(r)=\frac{1}{2\pi}e^{ik_xx}\psi_{k_z}(z)\phi_n(y-y_0),$$
$$ E_n(k_z)=\hbar \omega_c(n+1/2)+\frac{(\hbar k_z)^2}{2m}.$$
Here $n$ is a number of Landau level , $\omega_c = \frac{\displaystyle eH}{\displaystyle mc}$ is the cyclotron
frequency and $\phi_n(y)$ is the normalized harmonic-oscillator wave function with Landau state index. \\
As was mentioned above, applied  magnetic field  reduces the effective
dimensionality of charge from 3D to 1D, in other words electron gas in semiconductor can be considered as a set
of 1D gases and  every one dimensional gas  is specified by partial Fermi wave vector
$k_{\tt F}^n$. In the case of sufficiently weak fields $k_{\tt F}l_{\tt H} >>
1$, the partial Fermi wave vector can be defined as
$$
{k^n_{\tt F}}={k_{\tt F}}\sqrt{(1-\frac{2}{{(l_{\tt H}k_F)}^2}(n+1/2))}.
$$
Using  together with the expression for
polarization operator and we get an
equation for self-consistent potential in external magnetic field.
\begin{equation}
\Phi(\zeta)=-\frac{1}{2\pi} \int_0^{\infty} d\eta \int_{-\infty}^\infty dq_z
\frac{e^{iq_z\zeta}}{q_z^2\epsilon_{\tt H}(q_z)}K_{\tt H}(q_z,\eta)\Phi(\eta),
\end{equation}
where  $\epsilon_H(q_z) $ is  dielectric function in the
presence of magnetic field.
\begin{equation}
\epsilon_{\tt H}( q_z)=1+\sum_n\frac{4}{a_{\tt B} l^2_{\tt H}
q_z^3}\ln{\left|\frac{2{k_{\tt F}}^n+q_z}{2{k_{\tt F}}^n-q_z}\right|},
\end{equation}
$$
K_{\tt H}(q_z,\eta)=-\sum_n\frac{8}{a_{\tt B}{q_z}l^2_{\tt H}}({\tt Ci}|2{k_F}^n\eta+q_z\eta| -
{\tt Ci}|2{k_F}^n\eta-q_z\eta|).
$$
In the long-wave limit $q_z \ll k^n_{\tt F}$ the
expression for $\epsilon_{\tt H}$ can be transformed to
\begin{equation}
\epsilon_{\tt H}(q_z,0)=1-\frac{\sum_n {k_{\tt s}^n}^2}{q_{\tt z}^2},
\end{equation}
here ${k_{\tt s}^n}^2=\displaystyle {2}/{\pi a_{\tt B}l_{\tt H}^2k_{\tt
F}^n}$, $k^2_{s}=\sum_n {{k_s^n}^2}$
and summation is performed over all Landau levels where $k_{\tt s}^n \ll k_{\tt
F}^n$. It is evident that this condition  isn't valid for
all levels, however  using numerical modeling of (13) it is possible to demonstrate that  in a case of weak-magntetic fields
expression (14) is a qiute good approximation of (13), even though the summation is performed
over all occupied Landau levels.
It is easily seen that in the zero-magnetic-field limit
$k_s$ transforms to well-known expression for Thomas-Fermi vector.

In a similar to the zero-field case, in the long-wave limit the integral equation
(12) can be transformed to
\begin{equation}
\Phi''-({k_{\tt s}}^2-{\sum_n {k_s^n}^2}\cos{2{k_{\tt F}}^n\zeta})\Phi=0.
\end{equation}
\begin{equation}
\left. {\frac{d\Phi}{d\zeta}}\right|_{\zeta=0} = E \qquad
\left. {\frac{d\Phi}{d\zeta}}\right|_{\zeta=\infty} = 0
\end{equation}
This equation describes behaviour of electrostatic potential at $\zeta \sim \lambda_{\tt s}$,
here $\lambda_{s}$ is Thomas-Fermi length in magnetic field.
It is possible to show (using Poisson summation formula) that in the limit of zero magnetic field Eq.(15)
can be transformed to Eq.(9). \\
Similarly to the case $H=0$, long-range behaviour of $\Phi$ is conditioned by branch points in the dielectric function and at $\zeta \gg \lambda_{\tt TF}$
charge density oscillations have the following asymptotic
\begin{equation}
\varrho(\zeta) = \Phi(0)\sum_n
{
{k_{\small \rm s}^n} k_{\small \rm F}^n \frac {\cos{2k_{\small \rm F}^n\zeta}} {2k_{\small \rm
F}^n\zeta}.
}
\end{equation}
Every term in this sum decays as $1/\zeta$  and it is   typical for
one-dimensional systems. The same behaviour was obtained  in \cite{6}, however
in that article authors used exponential parametrization of
self-consistent potential, where unknown parameter
determined while functional minimization process.
In the case  $l_{\tt H}k_{\tt F} \gg 1$, summation in (17) can be
fulfilled using Poisson formula
\begin{equation}
\varrho(\zeta) \sim k_{\small \rm TF}k_{\small \rm F}\frac {\cos{2k_{\small \rm F}\zeta}} {(2k_{\small \rm F}\zeta)^2} +
 \frac{\sqrt2}{k_s a_{\tt B} \zeta l_{\tt H}}{\tt Re} \sum_{k=1}^{\infty}
(-1)^k \frac {e^{\displaystyle i\pi(k_{\small
\rm F}l_{\tt H})^2}}{\sqrt{k}}e^{\displaystyle -i\zeta^2/\pi k{l_{\small \rm H}}^2}
\end{equation}
The first term in this expression  corresponds to  density
oscillations without magnetic field and it is in an agreement with Eq.(11), the second term depends on magnetic field.\\
In the magnetic quantum  limit, when  only the lowest Landau
level is partially filled, we have
\begin{equation}
\varrho(\zeta)= \Phi(0)k_s {k_{\tt F}}\frac{\cos(2k_{\small \rm F}\zeta)}{2k_{\tt F}\zeta}
\end{equation}
here we need to mention that in quantum limit, $k_{\tt F}$ can't
be calculated using expression presented above.

\section{Electrostatic capacity of the barrier structure}
It is well-known that
the differential capacitance per unit area is $C=\displaystyle dQ/dV$
 where
$$Q=\displaystyle \int_{d/2}^{\infty}dz \varrho(z)$$ is a total charge over semiconductor
lead. Substituting distribution of self-consistent electron
density $\varrho(z)$ into the definition for electrostatic capacity we will
obtain an expression for $C$. In
the case of zero magnetic fields it expressed as
\begin{equation}
C=\frac{1}{4\pi}\frac{1}{(d+ 2\lambda_{\tt TF}+ 2\lambda_q)},
\end{equation}
where
$$
\lambda_{\tt q} = {r_{\tt s}}{\ln^{-1} \left(\frac{1}{\sqrt{r_{\tt s}}}\right)}\lambda_{\tt
TF}.
$$
Accounting of Friedel oscillations in electron density produce a
shift of charge center mass towards and away from the surface semiconductor-dielectric, what leads to decreasing capacity of the structure.
We wish to separate the classical term in the total capacity,
which depends on the barrier width and classical screening length
from quantum one which depends on long-range oscillations  of charge
density
$$
\frac{1}{C}= \frac{1}{C_{\tt g}}+ \frac{1}{C_{\tt s}} + \frac{1}{C_{\tt q}},
$$
here $C_{\tt g}$ is the geometrical capacity of the structure, $C_{\tt s}=\displaystyle \frac{1}{8\pi\lambda_{\tt TF}}$ is
contribution to capacity due to Thomas-Fermi screening in
semiconductor regions.
The expression for $C_{\tt s}$ coincides with result obtained within
semiclassical approach \cite{7}, which was based on solution of Boltzmann equation for electron distribution function.
$C_{\tt q}= \displaystyle \frac{1}{8\pi\lambda_q}$ is  quantum correction to tunnel capacity due to
long-range behaviour of charge density. As we mentioned in the introduction, in this study we assumed that $ r_{\tt
s}\ll 1$, so quantum contribution is smaller then the classical one
\begin{equation}
\frac{C_{\tt s}}{C_{\tt q}}=\frac{r_{\tt s}}{\ln{\displaystyle \frac{k_{\tt F}}{k_{\tt
TF}}}}\ll 1.
\end{equation}
In the case of finite  magnetic fields Friedel oscillations decay slower
than in the case $H=0$ and this fact brings to increasing to penetration length of  electric field
in semiconductor lead. For this case the expression for capacity
has the form
\begin{equation}
C_{\tt H}=\frac{1}{4\pi}\frac{1}{d+2\lambda_s+\displaystyle\sum_n{\frac{\displaystyle {k^n_s}^2}{\displaystyle k_s^2}\lambda_{\tt
F}^n}}.
\end{equation}
Under the condition $k_{\tt F}l_{\tt H} \gg 1$  we have
\begin{equation}
C_{\tt H} = \frac{1}{d + 2\lambda_{\tt TF}+ \lambda_{\tt q} + 2\lambda_{\tt TF}\displaystyle \frac{\sqrt{r_s}}{(k_{\tt F}l_{\tt
H})^2}\sum_{\tt k=1}\frac{(-1)^k}{\pi k}\sin{\pi k}(k_{\tt F}l_{\tt H})^2}
\end{equation}
Thomas-Fermi screening length also depends on $H$, but in a weak
magnetic fields it is approximately equal to $\lambda_{\tt TF}$.
\begin{equation}
\frac{C_{\tt s}}{C_{\tt q}(H)}\sim \frac {\sqrt {r_{\tt s}}}{(k_{\tt F}l_{\tt H})^2}
\end{equation}
As it is easy to see that $\displaystyle \frac{C_{\tt s}}{C_{\tt q}(H)}$ is still small however in difference from the previous case
this ratio
$\sim \sqrt{r_{\tt s}}$. Quantum contribution in the total capacity will
play more significant role with increasing of magnetic field.
\section {Conclusions}
Within RPA approximation expressions for self-consistent electrostatic potential are obtained. It is shown, that long-range electron
density oscillations lead to decreasing of structure capacity and
this effect plays more significant role with increasing external magnetic field.
\section{Acknowledgements}
This work has been  performed with support of the Russian Foundation for
Basic Research (Grant 03-02-16728)

\end{document}